\documentclass[prc,twocolumn,nofootinbib,superscriptaddress]{revtex4-1}

\usepackage{graphicx,amsmath,amssymb,bm,tabularx}
\usepackage{multirow,dcolumn}

\usepackage{xcolor}

\newcommand {\Fig} [1] {Fig.~\ref{#1}}
\newcommand {\Eq} [1] {Eq.~(\ref{#1})}

\newcolumntype{.}{D{.}{.}{2.3}}

\begin{document}

\title{Dispersion relations applied to double-folding potentials from chiral EFT}

\author{V.\ Durant}
\email[Email:~]{vdurant@uni-mainz.de}
\affiliation{Institut f\"ur Kernphysik, Johannes Gutenberg-Universit\"at Mainz, D-55099 Mainz, Germany}

\author{P.\ Capel}
\email[Email:~]{pcapel@uni-mainz.de}
\affiliation{Institut f\"ur Kernphysik, Johannes Gutenberg-Universit\"at Mainz, D-55099 Mainz, Germany}
\affiliation{Physique Nucl\'eaire et Physique Quantique (CP 229),Universit\'e libre de Bruxelles (ULB), B-1050 Brussels, Belgium}

\author{A.\ Schwenk}
\email[Email:~]{schwenk@physik.tu-darmstadt.de}
\affiliation{Institut f\"ur Kernphysik, Technische Universit\"at Darmstadt, 64289 Darmstadt, Germany}
\affiliation{ExtreMe Matter Institute EMMI, GSI Helmholtzzentrum f\"ur Schwerionenforschung GmbH, 64291 Darmstadt, Germany}
\affiliation{Max-Planck-Institut f\"ur Kernphysik, Saupfercheckweg 1, 69117 Heidelberg, Germany}

\begin{abstract}
We present a determination of optical potentials using the double-folding method based on chiral effective field theory nucleon-nucleon interactions at next-to-next-to-leading order combined with dispersion relations to constrain the imaginary part. This approach is benchmarked on $^{16}$O--$^{16}$O collisions, and extended to the $^{12}$C--$^{12}$C and $^{12}$C--$^{16}$O cases. Predictions derived from these potentials are compared to data for elastic scattering at energies up to 1000 MeV, as well as for fusion at low energy. Without adjusting parameters, excellent agreement with experiment is found. In addition, we study the sensitivity of the corresponding cross sections to the nucleon-nucleon interactions and nuclear densities used.
\end{abstract}

\maketitle

\section{Introduction}
One of the long-standing challenges in the study and description of nuclear reactions is the determination of the interaction between the colliding nuclei~\cite{Bran97heavyion}. Typically, these interactions are modeled using phenomenological 
potentials whose parameters are adjusted to reproduce elastic-scattering data, or obtained from inversion of scattering data~\cite{Leeb85PotInvdata}. These potentials reproduce experimental data precisely, but lack predictive power. 
Double-folding potentials (DFP) are nucleus-nucleus interactions constructed using the nucleonic densities of the reacting nuclei and the interaction between nucleons as input~\cite{Satc79Folding}. They present a way of determining potentials 
relevant for nuclear reactions based on more fundamental inputs: realistic nuclear densities and nucleon-nucleon ($NN$) interactions. Even though this framework provides more realistic potentials for the nucleon-nucleus interactions than for
the nucleus-nucleus case~\cite{Maha91eqpot}, interesting results have been obtained in such a way, e.g., by considering zero-range contact $NN$ interactions~\cite{Cham02SPdens,Pere09ImDFPot} or using a $G$-matrix approach, see, e.g., 
Refs.~\cite{Furu12OpPot,Mino15ChEFTCC} for recent work.
For modern nuclear forces, chiral effective field theory (EFT) has become the standard method for developing interactions rooted in the symmetries of quantum chromodynamics (see, e.g., Refs.~\cite{Epel09RMP,Mach11PR,Hamm13RMP} for reviews). Based on a power counting scheme, $NN$ interactions can be expressed as an expansion that starts at leading order (LO), followed by contributions at next-to-leading order (NLO) and next-to-next-to leading order (N$^2$LO), which leads to a systematic improvement of observables.
In a recent study~\cite{Dura17DFP}, we have explored the construction of double-folding potentials starting from local $NN$ interactions based on chiral EFT~\cite{Geze13QMCchi,Geze14long,Lynn14QMCln,Tews16QMCPNM,Lynn16QMC3N,Lynn17QMClight,Huth17Fierz}.

Double-folding potentials obtained with chiral EFT $NN$ interactions at the Hartree-Fock level are purely real. However, to properly reproduce scattering observables, an imaginary part needs to be added to simulate the absorption into non-elastic channels that can be open. 
In our previous work~\cite{Dura17DFP} it was simply assumed to be proportional to the real part using a proportionality constant $N_W=0.6-0.8$ motivated by Ref.~\cite{Pere09ImDFPot}.
The agreement of our results with elastic scattering data is good~\cite{Dura17DFP}, but we have observed a sensitivity to the choice of the imaginary part of the optical potential, especially at large scattering angles.
This sensitivity motivates the use of more refined descriptions of the imaginary part of the potential. Since the real and imaginary parts of the potential are related by dispersion relations~\cite{Fesh92Theo}, we apply them to derive the imaginary term of 
the optical potential, following Refs.~\cite{Carl89Dispersion,Bran90DisRel,Gonz01DisRel}, where these relations have been successfully used to constrain the energy-dependent terms of nucleus-nucleus potentials.

We focus on three systems: $^{16}$O--$^{16}$O, to compare with the work of Ref.~\cite{Dura17DFP}, $^{12}$C--$^{12}$C to extend this formalism to non-closed shell nuclei, and $^{12}$C--$^{16}$O to test the validity of this approach in
asymmetric collisions. As reaction observables, we consider the elastic-scattering cross sections and the astrophysical $S$ factors for the fusion at low energy.
In both cases, we test the sensitivity to the choice of the nuclear density, comparing between phenomenological two-parameter Fermi distributions~\cite{Cham02SPdens} and density profiles obtained from electron scattering~\cite{Devr87rhoel}. 

This paper is organized as follows: in Sec.~\ref{sec:formalism} we give a brief overview of the formalism for the double-folding potential and the reaction observables relevant for this study. In Sec.~\ref{sec:dispersion} we present the dispersion relations and 
their application to the elastic scattering of $^{16}$O--$^{16}$O, and $^{12}$C--$^{12}$C. We follow in Sec.~\ref{sec:density} with an analysis of the impact of different density profiles on the results for elastic scattering and astrophysical $S$ factors of 
the fusion for these two systems along with the asymmetric scattering of $^{12}$C--$^{16}$O. Finally, we summarize and give an outlook in Sec.~\ref{sec:sum}.

\section{Theoretical framework}
\label{sec:formalism}

\subsection{Double-folding potentials}

In the double-folding formalism, the nuclear part of the potential between nucleus 1 (with atomic and mass numbers $Z_1$ and $A_1$) and nucleus 2 (with $Z_2$ and $A_2$) can be constructed from a given $NN$ interaction $v$ by folding it over the corresponding
densities. The review of the formalism for the double-folding potential in this section follows Refs.~\cite{Furu12OpPot, Dura17DFP}. 
The resulting antisymmetrized potential can be written as a sum of the direct (D) and exchange (Ex) contributions: $V_\text{F}=V_\text{D}+V_\text{Ex}$. 

\begin{figure}[t]
\begin{center}
\includegraphics[width=0.75\columnwidth]{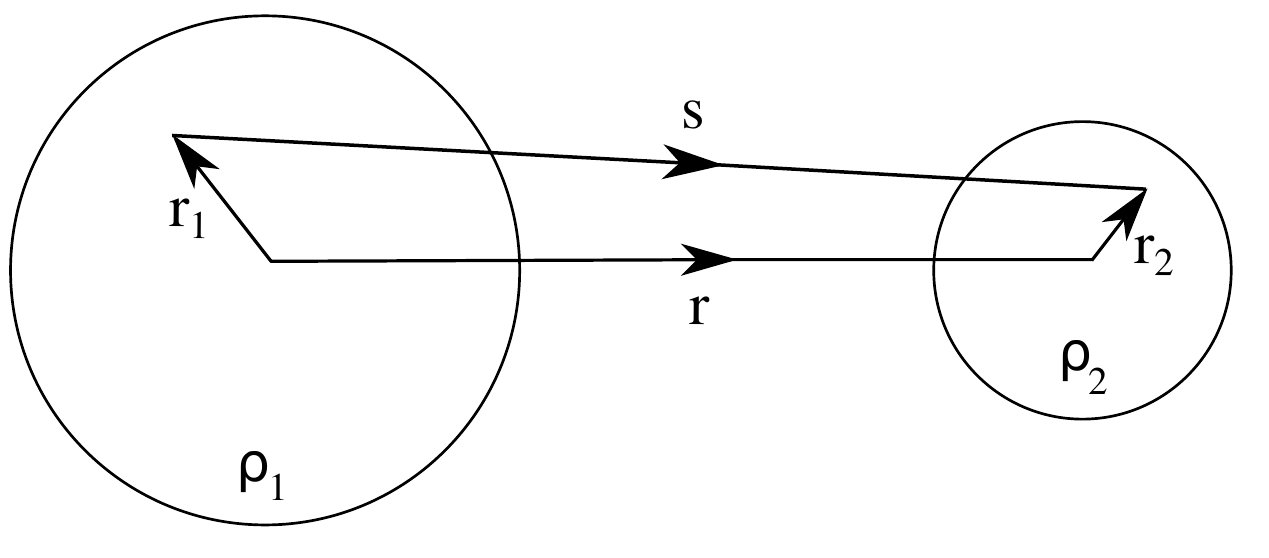}
\end{center}
\caption{\label{fig:coordinates}
Coordinates of the nuclei involved in the double-folding calculation [see Eqs.~(\ref{eq:direct}) and~(\ref{eq:exchange})].}
\end{figure}

Taking into account the coordinates of the geometry shown in \Fig{fig:coordinates}, the direct contribution to the double-folding potential is given by 
\begin{equation}
V_\text{D}(r) = \sum_{i,j = n,p} \iint \rho^i_1({\bf r}_1) \, v^{ij}_\text{D}({\bf s}) \, \rho^j_2({\bf r}_2)
\, d^3{\bf r}_1 d^3{\bf r}_2 \,,
\label{eq:direct}
\end{equation}
\noindent where ${\bf s}$ is given as in Fig.~\ref{fig:coordinates}, and $\rho_1^{i}$ and $\rho_2^{i}$ with $i=n,p$ are the neutron and proton density distributions of the colliding nuclei, respectively.

The exchange part of the potential reads
\begin{multline}
V_\text{Ex}(r,E_\text{cm}) = \sum_{i,j = n,p} \iint \rho^i_1({\bf r}_1,{\bf r}_1+{\bf s})
\, v^{ij}_\text{Ex}({\bf s}) \\
\times \rho^j_2({\bf r}_2,{\bf r}_2-{\bf s}) \exp \left[\frac{i{\bf k}(r)\cdot{\bf s}}{\mu/m_N}\right] \,
d^3{\bf r}_1 d^3{\bf r}_2 \,, \label{eq:exchange}
\end{multline}

\noindent where $\mu=m_NA_1A_2/(A_1+A_2)$ is the reduced mass of the colliding
nuclei (with $m_N$ the nucleon mass), and the integral is over the density matrices $\rho^i({\bf r},{\bf r} \pm {\bf s})$ of the nuclei. In this channel, there is an additional phase that renders the
double-folding potential dependent on the energy $E_\text{cm}$ in the center-of-mass reference frame. The momentum for the nucleus-nucleus relative motion ${\bf k}$ is related to $E_\text{cm}$, the nuclear part of the double-folding potential, 
and the double-folding Coulomb potential $V_\text{Coul}$ through
\begin{equation}
k^2(r)=2\mu \, \Bigl[ E_\text{cm} - V_\text{F}({r},E_\text{cm}) - V_\text{Coul}({r}) 
\Bigr] \,. \label{eq:k}
\end{equation}
As a result, $V_\text{Ex}$ has to be determined self-consistently. The density matrices entering in \Eq{eq:exchange} are approximated using the density matrix expansion restricted to its leading term~\cite{Nege72DME1} (see also the discussion in Sec.~II of 
Ref.~\cite{Dura17DFP}).

For the calculation of the double-folding potentials used in this work, we include only two-body forces. We take the local chiral $NN$ interactions regulated with cutoffs $R_0=1.2$, $1.4$ and $1.6$~fm presented in 
Ref.~\cite{Dura17DFP} following Refs.~\cite{Geze13QMCchi,Geze14long}.
It is interesting to note that for doubly closed-shell nuclei, like $^{16}$O, the $NN$ interaction used in the the double-folding formalism receives contributions only from the central parts of nuclear forces. In the case of open-shell nuclei,
$^{12}$C in this study, the spin-orbit and tensor contributions to $NN$ interactions need to be also taken into account. 

\subsection{Reaction observables}

As in Ref.~\cite{Dura17DFP}, to test the validity of the double-folding method, we focus on two types of reactions: fusion and elastic scattering.

In the case of nuclear fusion involving collisions of light or medium-mass nuclei at energies in a range that goes from below to slightly above the Coulomb barrier, one usually assumes that the nuclear potential is purely real, since its 
imaginary part is well inside the range of the effective potential. 
For light systems, the fusion barrier is located before the neck formation, which justifies the use of the double-folding procedure. 
Then, the effective potential $V_\text{eff}$ is formed by the real double-folding potential~\cite{Hagi12IWBC}, the Coulomb potential between the nuclei and a centrifugal barrier that depends on the orbital angular momentum $l$.

At low energy, the projectile and target can penetrate the Coulomb and centrifugal barriers thanks to the tunnel effect. Once the nucleus is within the barrier, its probability to get out is so low that it can be neglected. This situation 
is described by the incoming-wave boundary condition (IWBC)~\cite{Land84IWBC}, under which the fusion cross section can be obtained from the probability to tunnel through the barrier in each of the partial waves~\cite{Hagi12IWBC}. 
The fusion cross sections are determined using the code CCFULL~\cite{Hagi99IWBCcode}, in which we have included 
the effects of the symmetrization of the wave function needed when the fusing
nuclei are identical spinless bosons.

The elastic scattering of medium to heavy nuclei can be described within the optical model. The nuclear part of the interaction between the colliding nuclei is described by a complex potential, whose imaginary part accounts for the probability
that the system leaves the elastic channel. In Ref.~\cite{Dura17DFP} we have assumed the imaginary part of the potential to be proportional to its real part with a proportionality constant $N_W=0.6-0.8$. It was seen then that elastic-scattering calculations 
are sensitive to the choice of the imaginary part of the potential, and that this description needs to be refined. 

\section{Dispersion relations}
\label{sec:dispersion}

\subsection{Formalism}

Following Feshbach's formalism~\cite{Fesh92Theo} a local complex optical potential $U$ between two nuclei can be written as the sum of three contributions: a real term independent of the energy, a real term dependent on the energy and an imaginary term. 
In our case,
\begin{equation}
U_F(r,E_\text{cm})=V_\text{D}(r)+V_\text{Ex}(r,E_\text{cm})+iW(r,E_\text{cm})\,.
\end{equation} 
The dispersion relation holds between the energy-dependent real and imaginary parts, and reads~\cite{Carl89Dispersion,Gonz01DisRel}
\begin{equation}
W(r,E_\text{cm})=\frac{1}{\pi}\mathcal{P}\int_{-\infty}^{+\infty} dE'\frac{V_\text{Ex}(r,E')}{E'-E_\text{cm}}\,,
\end{equation}
\noindent where $\mathcal{P}$ represents the principal value integral.

In the case of our double-folding potentials, $V_\text{Ex}$ has a nearly identical radial dependence at all energies for both systems, and only its depth varies with the energy. We can then write the exchange part of the potential as 
a purely radial part $f_\text{Ex}$ and a potential depth $V^0_\text{Ex}$ that carries the energy dependence
\begin{equation}
V_\text{Ex}(r,E)=V^0_\text{Ex}(E)f_\text{Ex}(r)\,,
\end{equation}

\noindent which leads to

\begin{equation}
W(r,E_\text{cm})=\frac{ f_\text{Ex}(r)}{\pi}\mathcal{P}\int_{-\infty}^{+\infty} dE'\frac{V^0_\text{Ex}(E')}{E'-E_\text{cm}}\,.
\label{eq:W_disp}
\end{equation}

Because the integral in Eq.~(\ref{eq:W_disp}) requires the depth of $V_\text{Ex}$ at negative energies, we set $V^0_\text{Ex}(E'<0)=V^0_\text{Ex}(E'=0)$. We have checked that setting $V^0_\text{Ex}(E'<0)=0$ has no impact for energies higher 
than $E_\text{cm}\approx 30$ MeV, which is below the range of interest in this study.

\subsection{Results at N$^2$LO}

\begin{figure}[]
\begin{center}
\includegraphics[width=0.99\columnwidth]{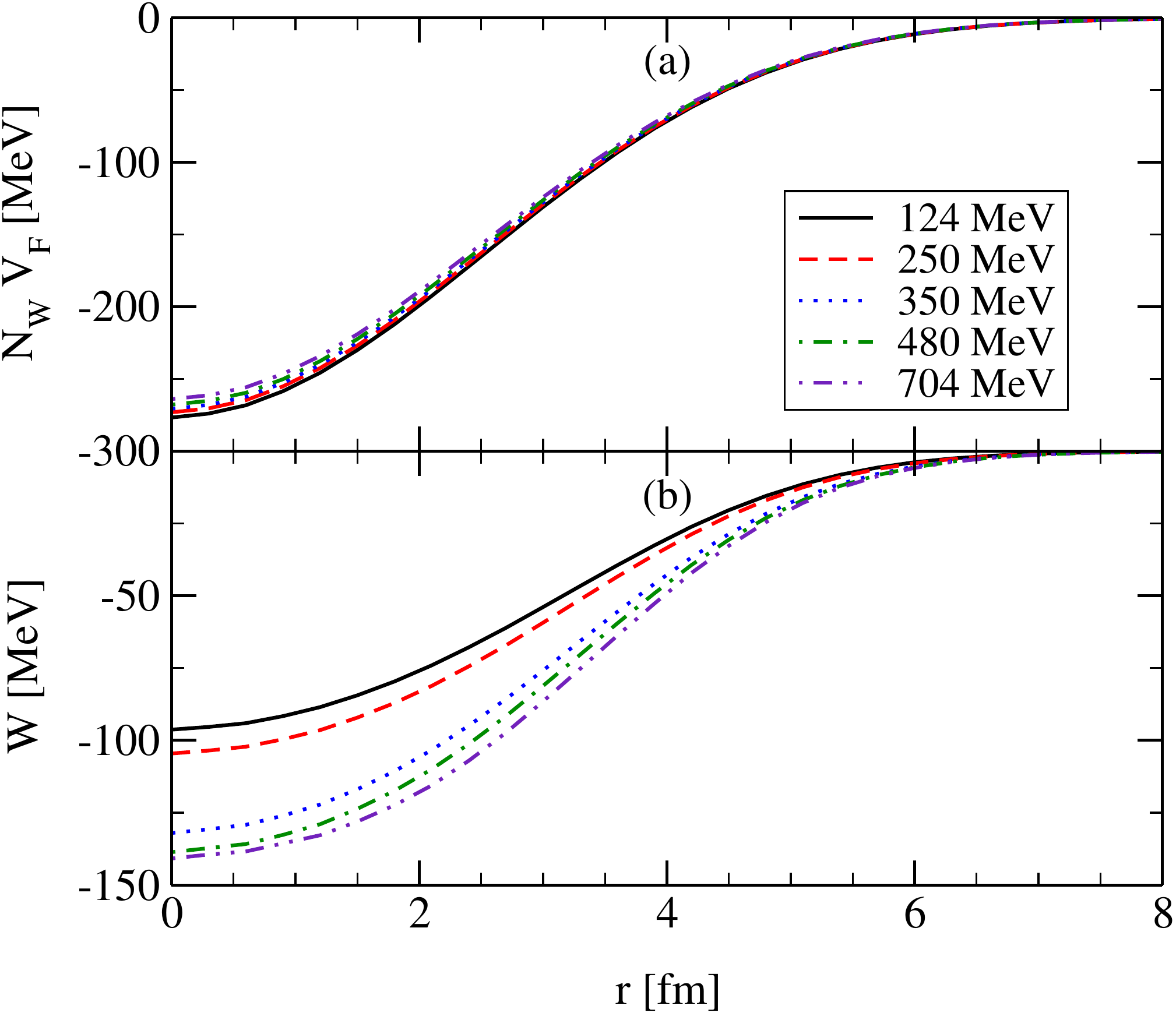}
\caption{Imaginary part of the double-folding potential for the $^{16}$O--$^{16}$O system: (a) proportional to the real double-folding potential with $N_W=0.6$ and (b) obtained through dispersion relations. The shown results are based on the local chiral EFT interaction at N$^2$LO with $R_0=1.4$~fm. The nucleonic densities were taken as two-parameter Fermi density distributions~\cite{Cham02SPdens}.}
\label{fig:potentials_16O}
\end{center}
\end{figure}

\begin{figure*}[htb]
\begin{center}
\includegraphics[width=1.8\columnwidth]{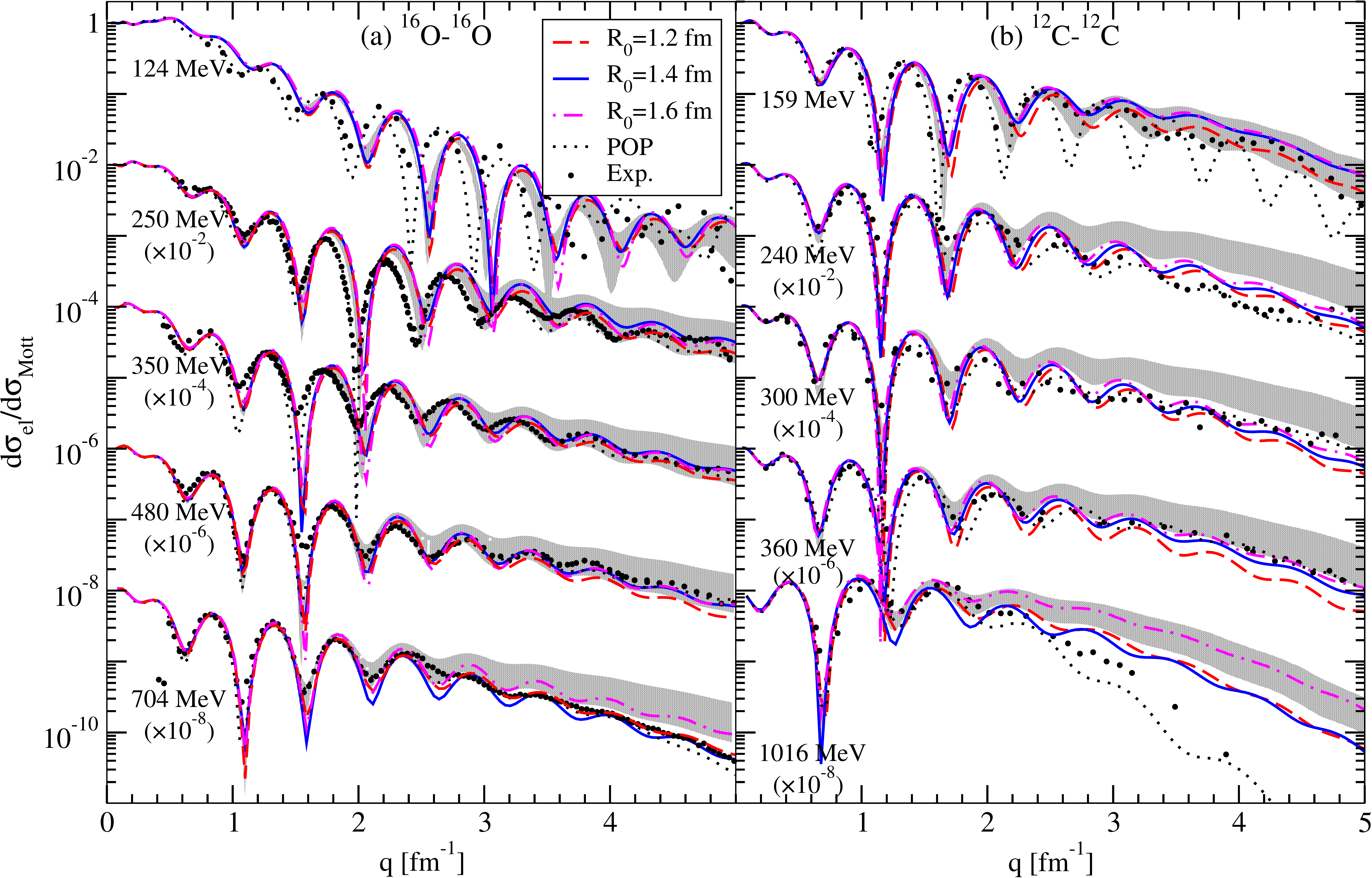}
%\hspace{-0.24cm}
%\includegraphics[width=0.83\columnwidth]{CS_12C12C_disp_q.pdf}
\caption{Cross section for elastic scattering of (a) $^{16}$O--$^{16}$O and (b) $^{12}$C--$^{12}$C as a function of momentum transfer~$q$ for different laboratory energies (normalized to the Mott cross section). Results are shown using $NN$ potentials
with $R_0=1.2$ (red), 1.4 (blue), and 1.6 fm (magenta) and dispersion relations to calculate the imaginary part of the double-folding potential. The shaded grey area shows results obtained in Ref.~\cite{Dura17DFP} using the simple $N_W=0.6-0.8$ prescription. 
The experimental data is shown as black circles and was taken from Refs.~\cite{Bohl93expel,Kond96expel,Bart96expel,Nuof98expel,Nico99expel,Khoa0016OEl,Kubo83C12,Bohl82C12,Buen81C12,Buen84C12,
Demy10C12el240}.
For comparison, the dotted lines show results obtained with phenomenological optical potentials (POP)~\cite{Khoa0016OEl,Kubo83C12,Bohl82C12,Buen81C12,Buen84C12,Demy10C12el240}. }
\label{fig:CS_2pF_q}
\end{center}
\end{figure*}

We first present potentials from double-folding interactions calculated with two-parameter Fermi density distributions~\cite{Cham02SPdens}. 
%fitted by the S\~ao Paulo group. 
As an example, Fig.~\ref{fig:potentials_16O} shows the imaginary part $W$ for $^{16}$O--$^{16}$O scattering at different laboratory energies, obtained at N$^2$LO with $R_0=1.4$~fm. Similar results are obtained with different cutoffs $R_0$ and at different chiral orders, for this system as well as for the other systems ($^{12}$C--$^{12}$C and $^{12}$C--$^{16}$O). In panel (a) we show the results assuming that the imaginary part is proportional to the real double-folding potential $V_\text{F}$, setting the proportionality constant to $N_W=0.6$. Panel (b) shows the new results applying dispersion relations, given by Eq.~(\ref{eq:W_disp}). From this comparison, it is clear that the imaginary part of the potential obtained with the dispersion relations exhibits a stronger energy dependence than if it is simply assumed proportional to the real part $V_{\rm F}$. Interestingly, dispersion relations lead to a reversed order in the potential depth compared to the real part, with higher energies leading to larger imaginary terms. This seems more reasonable, since we expect more open channels at high energy, and hence more absorption from the elastic channel. Note also that as $W$ is built exclusively from the weaker exchange part of the folding potential, the depth of the imaginary potential is significantly reduced compared our previous assumption.

The cross sections obtained in this work, using dispersion relations instead of the $N_W$ factor, show the same systematic behavior from chiral EFT that was seen in our previous work (see Figs.~4 and~5 of Ref.~\cite{Dura17DFP}) for both 
$^{16}$O--$^{16}$O and $^{12}$C--$^{12}$C. 
The only exception to this systematics appears for $^{16}$O--$^{16}$O at intermediate and high energies (480 MeV and above), where the exchange potential at leading order is repulsive, resulting in a non-physical elastic scattering cross section. 
This issue is resolved at NLO and N$^2$LO, where the behavior is consistent with that at lower energies. For this reason, we show only results at N$^2$LO.

Figure~\ref{fig:CS_2pF_q} shows the corresponding elastic-scattering cross sections 
(ratio to the Mott cross section) at different energies plotted as a function of the momentum transfer $q$. It can be seen that there is good agreement with experimental data for (a) $^{16}$O--$^{16}$O~\cite{Bohl93expel,Kond96expel,Bart96expel,Nuof98expel,Nico99expel,Khoa0016OEl} and (b) $^{12}$C--$^{12}$C~\cite{Kubo83C12,Bohl82C12,Buen81C12,Buen84C12,Demy10C12el240}. We also show results using different $NN$ cutoffs $R_0=1.2$, 1.4 and 1.6 fm 
(red, blue and magenta lines). It is clear that the application of dispersion relations leads to an improved description of experimental data compared to the scaling of the real part studied in Ref.~\cite{Dura17DFP} (shown by the shaded area). 
These new results show less uncertainty related to the description of the imaginary part, as it can be seen by the small dependence on the $NN$ cutoff $R_0$ at all energies for both systems. Albeit small, this sensitivity to $R_0$ increases at 
large momentum transfer, suggesting that the data are more sensitive to short-range physics at larger angles. The agreement with experiment is comparable at small and large momentum transfers, in contrast to the results found in Ref.~\cite{Dura17DFP}, 
where the results at large $q$ did not agree with the data (see shaded area). This might be due to the more realistic change in magnitude of the absorptive term given by dispersion relations. The only exception is found at high energy ($E_\text{lab}=1016$ MeV) for $^{12}$C--$^{12}$C, where the comparison deteriorates when the momentum increases. Our results are also 
in good agreement with cross sections obtained with phenomenological optical potentials (POP) for $^{16}$O--$^{16}$O~\cite{Khoa0016OEl} and $^{12}$C--$^{12}$C~\cite{Kubo83C12,Bohl82C12,Buen81C12,Buen84C12,Demy10C12el240} (shown by the dotted black lines). 

For low collision energies there is a shift in the oscillations of our results towards larger momentum transfers, especially in the case of $^{16}$O--$^{16}$O. This shift was already seen in Ref.~\cite{Dura17DFP} and suggests that the description of our 
potentials at low energies needs still more refinement.

\section{Impact of the density}
\label{sec:density}

\subsection{Density profiles}

\begin{figure}[t]
\begin{center}
\includegraphics[width=0.99\columnwidth]{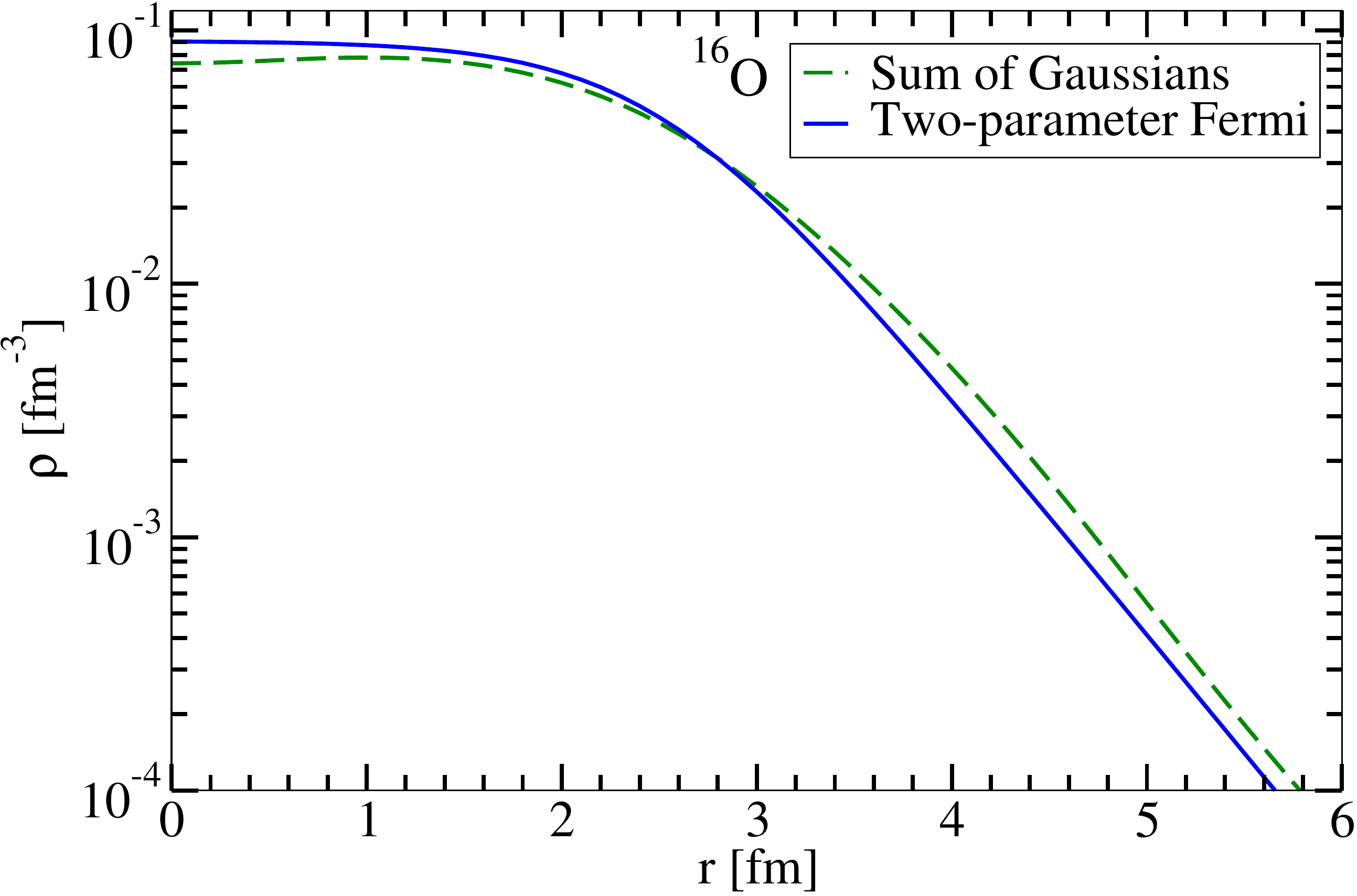}
\caption{Proton density of $^{16}$O using a two-parameter Fermi distribution~\cite{Cham02SPdens} (blue) and the sum of Gaussians parametrization from electron scattering~\cite{Devr87rhoel} (green dashed).}
\label{fig:dens_16O}
\end{center}
\end{figure}

To study the sensitivity of the reaction observables to the nuclear density, we consider different realistic densities. We compare the results obtained with the two-parameter Fermi densities of Ref.~\cite{Cham02SPdens} (see Sec.~\ref{sec:dispersion}) 
to calculations obtained using densities obtained from electron-scattering experiments listed in Ref.~\cite{Devr87rhoel}, for which we use two parametrizations based on a Fourier-Bessel as well as a sum of Gaussians. For the nuclei involved in this study, these two density profiles give almost indistinguishable cross sections for elastic scattering and fusion at different energies. For this reason, we show results using only the sum of Gaussians parametrization. Since we are describing light nuclei with the same number of protons and neutrons, we assume $\rho^n=\rho^p$ (see Fig.~\ref{fig:dens_16O} for a comparison between two-parameter Fermi and sum of Gaussians $^{16}$O proton densities).

In general, a potential obtained through a double-folding procedure depends on the choice of the nuclear densities. We have observed that, for the systems studied in this work, the exchange part of the double-folding potential is more affected by the density 
choice than its direct part. Since $V_\text{D}$ is one order of magnitude larger than $V_\text{Ex}$ (see Fig. 3 of Ref.~\cite{Dura17DFP}), the impact of the density in the cross sections is small when the imaginary part is taken to be proportional to the real 
potential.
However, if we apply the dispersion relations to describe the imaginary part of the interaction, $W$ will have the radial shape of $V_\text{Ex}$ [see Eq.~(\ref{eq:W_disp})]. In this case, optical potentials obtained with different density profiles will have 
imaginary parts with different shapes. Since the imaginary potential has a significant impact on the cross sections, it is interesting to study how a different density parametrization influences the results for elastic scattering. 

\subsection{Elastic scattering}
\label{sec:CS_rho}

\begin{figure*}[htb]
\begin{center}
\includegraphics[width=1.8\columnwidth]{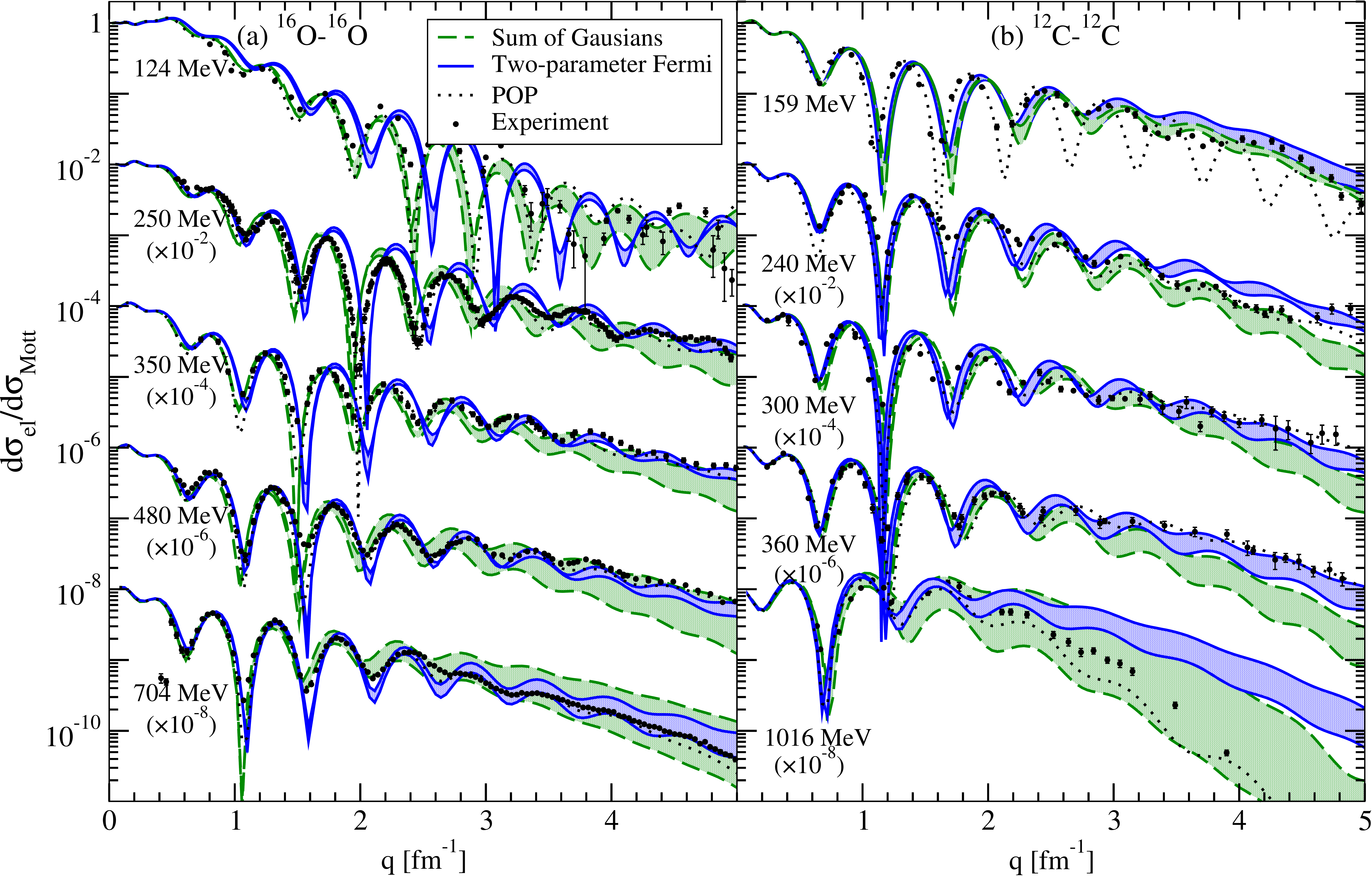}
\caption{Influence of the nucleonic density on the calculated cross sections for elastic scattering of (a) $^{16}$O--$^{16}$O and (b) $^{12}$C--$^{12}$C. The ratio to the Mott cross section is shown as a function of the momentum transfer~$q$ for
different laboratory energies. Results are calculated using a two-parameter Fermi distribution~\cite{Cham02SPdens} (blue) and sum of Gaussians density parametrization~\cite{Devr87rhoel} (green dashed). The shaded areas illustrate the 
sensitivity to $R_0 = 1.2$--$1.6$~fm at N$^2$LO. 
The experimental data is shown as black circles and was taken from Refs.~\cite{Bohl93expel,Kond96expel,Bart96expel,Nuof98expel,Nico99expel,Khoa0016OEl,Kubo83C12,Bohl82C12,Buen81C12,Buen84C12,
Demy10C12el240}.
For comparison, we also show phenomenological optical-potential (POP) results as dotted lines~\cite{Khoa0016OEl,Kubo83C12,Bohl82C12,Buen81C12,Buen84C12,Demy10C12el240}. }
\label{fig:CS_SG_q}
\end{center}
\end{figure*}

\begin{figure}[htb]
\begin{center}
\includegraphics[width=\columnwidth]{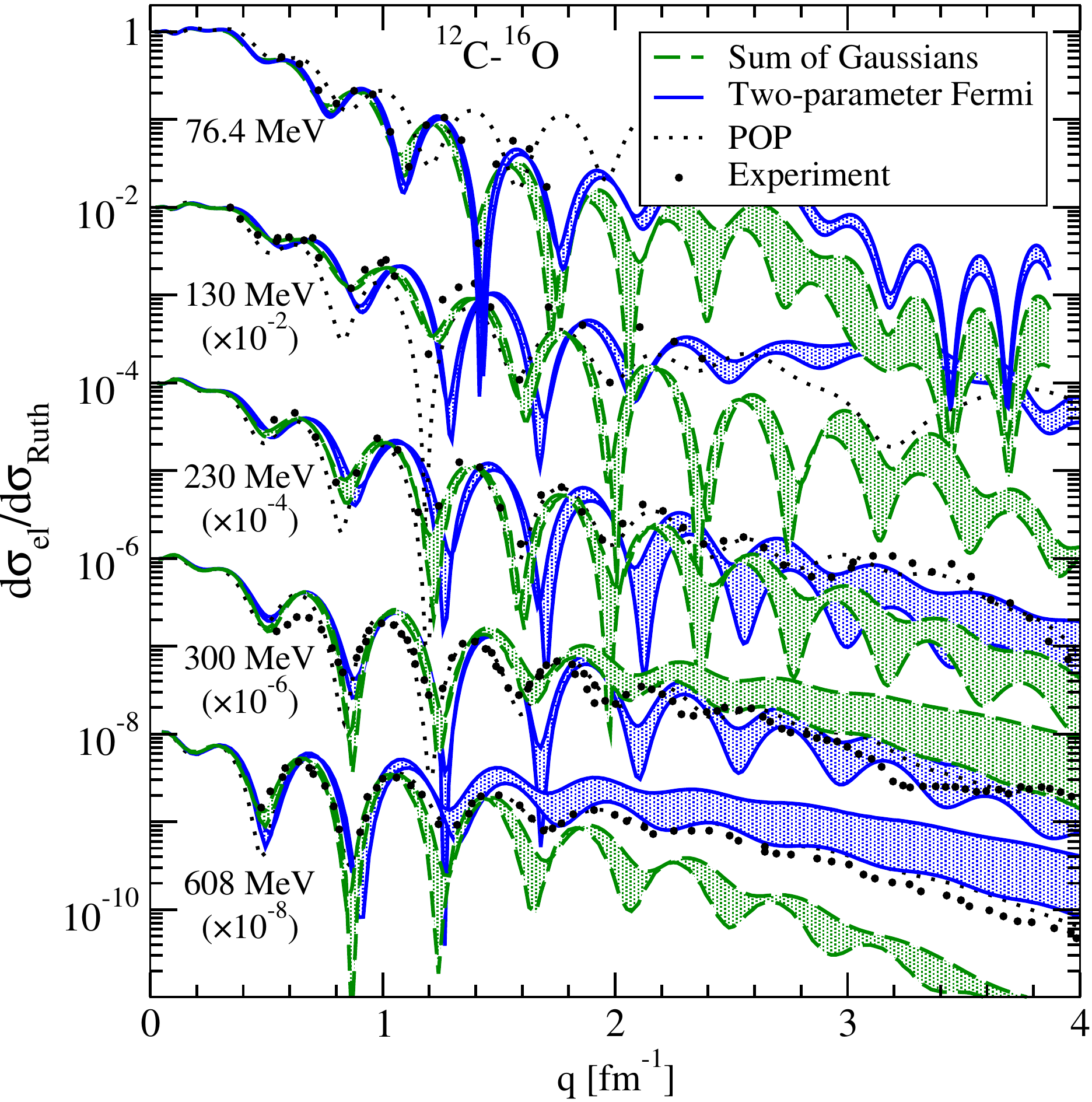}
\caption{Same as Fig.~\ref{fig:CS_SG_q} for elastic-scattering of the asymmetric system $^{12}$C--$^{16}$O.
The experimental data (black circles) and the phenomenological optical potentials (black dotted) are taken from Refs.~\cite{Moto79COel,Ikez86COel,Bran86COel,Oglo00COel,
Bran01COel}.}
\label{fig:CS_SG_q_ant}
\end{center}
\end{figure}

Figure~\ref{fig:CS_SG_q} shows the results for elastic-scattering cross sections using two-parameter Fermi density profiles~\cite{Cham02SPdens} (blue lines) and densities derived from electron-scattering experiments~\cite{Devr87rhoel} (green dashed lines). 
The bands give the sensitivity to $R_0 = 1.2$--$1.6$~fm at N$^2$LO. It can be clearly seen that using realistic densities combined with dispersion relations significantly improves the results. The fact that the magnitude of the exchange potential decreases 
as the collision energy increases (see Fig. 3 of Ref.~\cite{Dura17DFP}) also explains why the improvement is more significant at low energies. At small momentum transfer, where our model is most reliable, the reproduction of experimental data is significantly 
enhanced. In the case of $^{16}$O--$^{16}$O collisions, the improvement of the aforementioned shift of the minima at low energies is remarkable. There remain some discrepancies between the data and our results at large momentum transfers, which are mainly 
determined by short-range physics.

Figure~\ref{fig:CS_SG_q_ant} shows the results for the asymmetric $^{12}$C--$^{16}$O scattering. In general, we observe the same kind of trend as that described for Fig.~\ref{fig:CS_SG_q}, and we find good reproduction of the experimental data.
However, it can be observed that at intermediate collision energies ($E_\text{lab}=300$ MeV) the magnitude of the cross section at low momentum transfer is larger than the experimental data for both density profiles. This is a feature also presented by the 
phenomenological optical potential parametrization of Ref.~\cite{Bran86COel}, as well as in modern parametrizations, such as Ref.~\cite{Ngu1812C16O}, in which coupled channel effects are included in order to model $\alpha$-cluster transfers. 
In general, we see that for this system our results describe experimental data for large momentum transfer less accurately than what can be observed for symmetric collisions. This is an indication that short-range effects are more important for asymmetric 
scattering. 
Since in this case partial waves with odd $l$ also contribute to the cross section, we expect more channels to be open. There could be effects from excitations given by structure and in-medium effects that are not included in the double-folding 
model applied here. This should be further investigated with the study of different asymmetric collisions.

It is important to note that the results obtained with two-parameter Fermi densities exhibit a shift in the minima between $R_0=1.2$ and 1.6 fm, even though the corresponding bands are, in general, narrow. The large dependence on $R_0$ for the double-folding potentials calculated with realistic densities shown by the green bands of Fig.~\ref{fig:CS_SG_q} indicates that there is more dependence on the absorptive term of the potentials. We have observed that the radial part of $V_\text{Ex}$ is more dependent on $R_0$ when using densities from electron scattering. Through the application of the dispersive relations [Eq.~(\ref{eq:W_disp})], 
this leads to a larger dependence on $R_0$ also for the imaginary potential, which controls the description of the absorptive channels. 

It can be seen from Figs.~\ref{fig:CS_SG_q} and~\ref{fig:CS_SG_q_ant} that, both for symmetric and asymmetric collisions, our results show excellent agreement with experimental data. We remind the reader that all observables are obtained without adjusting parameters in the nucleus-nucleus potentials.

\subsection{Fusion}
\label{sec:fusion}

At low energy, the fusion process is strongly hindered by the Coulomb repulsion, which makes the cross sections plummet when $E_\text{cm}$ decreases. This hindrance is well accounted for by the Gamow factor, which is usually factorized out of the cross section
to define the astrophysical $S$ factor
\begin{equation}
S(E_\text{cm})=E_\text{cm} \, e^{2\pi\eta} \, \sigma_{\text{fus}}(E_\text{cm}) \,,
\end{equation}
where the Sommerfeld parameter is given by $\eta = Z_1 Z_2 e^2/(4 \pi
\varepsilon_0 v)$, with $v$ the asymptotic relative velocity between the two
nuclei.

\begin{figure}[]
\begin{center}
\includegraphics[width=\columnwidth]{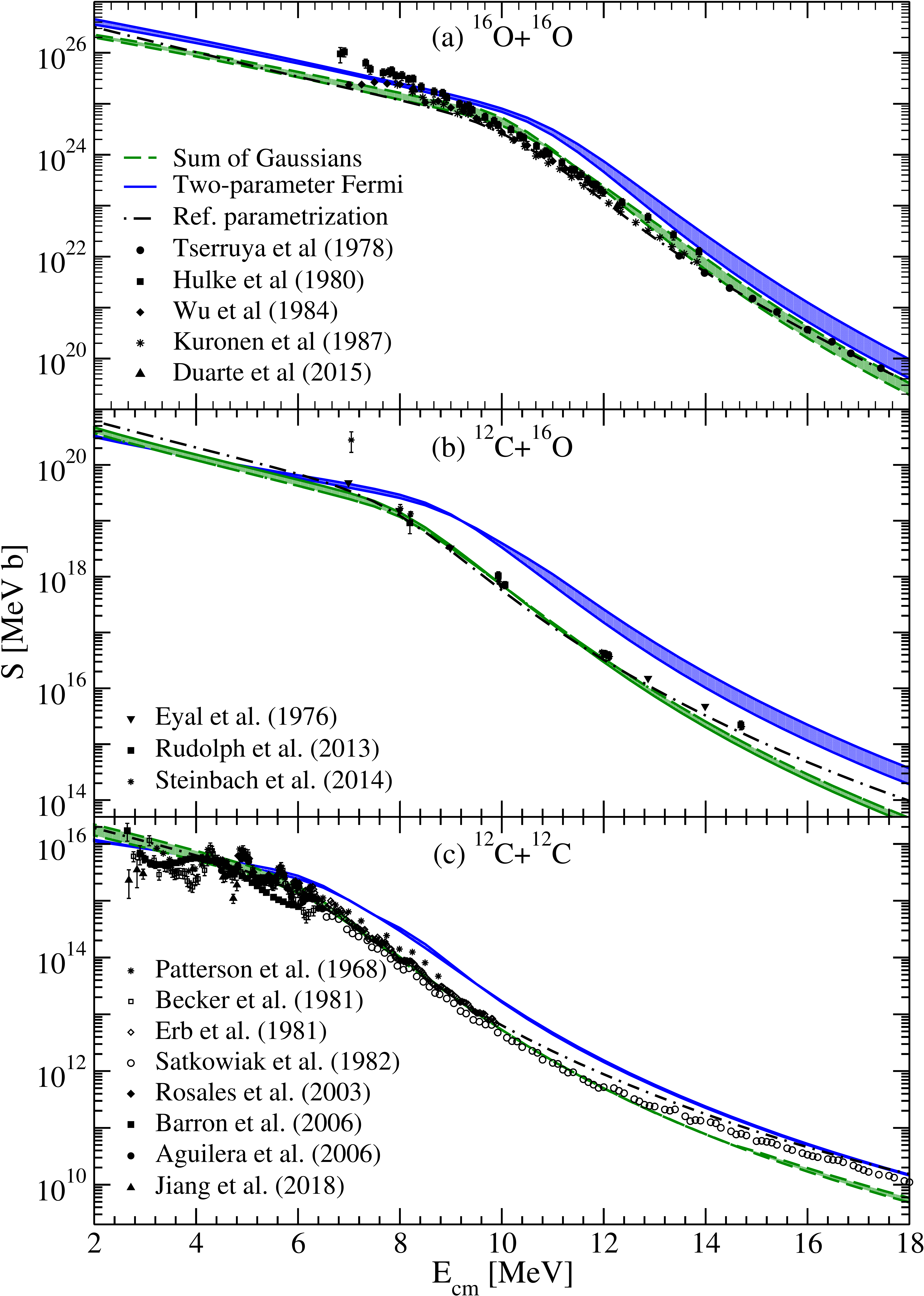}
\caption{Astrophysical $S$ factor for the fusion of (a) $^{16}$O$+^{16}$O, (b) $^{12}$C$+^{16}$O, and (c) $^{12}$C$+^{12}$C as a function of energy $E_\text{cm}$ in the center-of-mass system. Results are obtained at N$^2$LO using the two-parameter Fermi 
distribution~\cite{Cham02SPdens} (blue) and sum of Gaussians from~\cite{Devr87rhoel} (green dashed). The shaded areas illustrate the sensitivity to $R_0 = 1.2$--$1.6$~fm. The results of the parametrization from Ref.~\cite{Yako10Sfct} are displayed as black
lines. The black symbols depict experimental data from Ref.~\cite{Tser78expfus,Hulk80expfus,Wu84expfus,Kuro87expfus,
Duar15expfus},
~\cite{Patt6912CFus,Beck8112CFus,Erb8112CFus,Satk8212CFus,
Rosa0312CFus,Barr0612CFus,Agui0612CFus},
and~\cite{Eyal76COfus,Beck84COfus,Rudo12COfus,Stei14COfus}
.} 
\label{fig:S_fac}
\end{center}
\end{figure}

For the $S$ factors shown in this work, $V_{\text{Ex}}$ is taken at the center of the considered energy range, $E_\text{cm} = 12$~MeV, since the energy dependence in this range is negligible.
All results shown here are obtained using chiral $NN$ interactions at N$^2$LO. For a discussion of the order-by-order behavior of the $^{16}$O+$^{16}$O $S$ factor, we refer to Sec.V of Ref.~\cite{Dura17DFP}. The same kind of behavior is also observed for
$^{12}$C$+^{16}$O and $^{12}$C$+^{12}$C.
In the code used for the computation of the fusion cross section, we approximate the Coulomb interaction by a sphere-sphere potential of radius $R_C=r_C A_1^{1/3}+r_C A_2^{1/3}$, with $r_C=1.79$~fm~\cite{Baye82SphCoul}. We do not expect this change from the double-folding Coulomb term used in \Eq{eq:k} to affect significantly our results. 

Figure~\ref{fig:S_fac} shows the $S$ factor for (a) $^{16}$O$+^{16}$O, (b) $^{12}$C$+^{16}$O, and (c) $^{12}$C$+^{12}$C fusion obtained with two-parameter Fermi densities~\cite{Cham02SPdens} (blue lines) and profiles from electron scattering 
experiments~\cite{Devr87rhoel} (green dashed lines). The bands give the sensitivity to $R_0 = 1.2$--$1.6$~fm.
It can be clearly seen that the nuclear density plays a significant role in the fusion cross section, having much more impact than the sensitivity to the short-range physics. The results obtained with electron-scattering densities show excellent agreement with experimental 
data~\cite{Tser78expfus,Hulk80expfus,Wu84expfus,Kuro87expfus,Duar15expfus, Patt6912CFus,Beck8112CFus,Erb8112CFus,Satk8212CFus,Rosa0312CFus,Barr0612CFus,Agui0612CFus,Eyal76COfus,Beck84COfus,Rudo12COfus,
Stei14COfus}, in contrast to the $S$ factors calculated with two-parameter Fermi densities, which describe the results only qualitatively. We find that the use of realistic densities is crucial to reproduce the data. The data has been plotted with error bars, 
when available, except for the case of the measurement of Erb \textit{et al.}~\cite{Erb8112CFus} for $^{12}$C$+^{12}$C (empty diamonds), where the $\pm10\%$ error at all energies is not shown for readability of the figure.

In the case of $^{12}$C$+^{12}$C, note that the barrier penetration model applied in this work does not include the possibility of resonant states or hindrance of the fusion process, mechanisms that are reflected in the data from Refs.~\cite{Agui0612CFus} 
(filled circles) and~\cite{Jian18FusCa} (triangles), respectively. We also show theoretical parametrizations of the $S$ factors obtained by the S\~ao Paulo group~\cite{Yako10Sfct} (dash-dotted black lines), which are confirmed by our results for the $S$-factors
of these systems when using realistic densities. 

\section{Conclusions and Outlook}
\label{sec:sum}
We have presented the derivation of optical potentials using the double-folding method with local chiral EFT $NN$ potentials~\cite{Geze13QMCchi,Geze14long} and realistic nucleonic densities combined with dispersion relations to determine the imaginary part.
The application of these relations helps constraining efficiently the imaginary term of the nucleus-nucleus interactions, which are generated with no fitting or scaling parameter.
The use of these potentials gives excellent reproduction of elastic-scattering data at several energies for the collision of closed and non-closed shell nuclei as well as scattering of non-identical nuclei, as it was shown for the cases 
$^{16}$O--$^{16}$O, $^{12}$C--$^{12}$C, and $^{12}$C--$^{16}$O, respectively. 

The use of dispersion relations to calculate the imaginary potential leads to a better reproduction of data than in our previous study~\cite{Dura17DFP}, in which the imaginary part was simply assumed to be proportional to the real double-folding potential.
Moreover, adopting realistic density profiles from electron scattering~\cite{Devr87rhoel} instead of two-parameter Fermi parametrizations~\cite{Cham02SPdens} in the folding procedure gives significant improvement in the comparison with experiment,
both for elastic scattering and fusion.

We consider the use of realistic densities profiles combined with dispersive relations a necessary first step towards a better description of the imaginary part of nucleus-nucleus potentials. There are several avenues for improvement,
both at the level of the input interactions and the many-body folding method. First of all our investigation should be extended to other non-symmetric systems and to more exotic nuclei in the future. 
Also, it would be interesting to study the impact of going beyond leading order in the density matrix expansion. It is also necessary to determine the impact of a calculation beyond
Hartree-Fock and the nonlocal contributions that
would arise (see, e.g., Refs.~\cite{Fesh58,Fesh62}). Finally, the role of three-nucleon interactions needs to be investigated in this approach, as they also enter at N$^2$LO.

As a general feature of our results, we can conclude that there is excellent agreement between our calculations of observables and experimental data. It is important to remember that there is no fitting or scaling parameter in the nucleus-nucleus
potential.
These results hint strongly towards the interest of studying the impact of using density profiles based also on chiral EFT interactions to analyse the results within a fully consistent model that would bridge reactions and structure.

\section*{Acknowledgments}
We thank A.\ B.\ Balantekin for useful discussions and L.\ Gasques for providing the data on $^{12}$C+$^{12}$C fusion. We also thank the International Atomic Energy Agency that provided the experimental data through their web page \href{www-nds.iaea.org}{www-nds.iaea.org}. 
This work was supported by the PRISMA+ (Precision Physics, Fundamental Interactions and Structure of Matter) Cluster of Excellence, the European Union's Horizon 2020 research and innovation programme under Grant Agreement No. 654002, and 
Deutsche Forschungsgemein-schaft (DFG, German Research Foundation) -- Projekt-ID 279384907 -- SFB 1245 and Projekt-ID 204404729 -- SFB 1044.

\bibliography{references}

\end{document}